\documentclass[
    ,final            
  ]
  {aipproc}

\layoutstyle{8x11double}

\newcommand{\bea}{\begin{eqnarray}}
\newcommand{\eea}{\end{eqnarray}}
\newcommand{\be}{\begin{equation}}
\newcommand{\ee}{\end{equation}}
\newcommand{\nn}{\nonumber}
\newcommand{\lnum}{{\cal L}}
\newcommand{\bnum}{{\cal B}}

\begin{document}

\title{Lightest $U$-parity Particle (LUP): \\
a hidden sector dark matter candidate}



\author{Hye-Sung Lee}{
  address={Institute for Fundamental Theory, University of Florida, Gainesville, FL 32611, 
USA}
}




\begin{abstract}
We introduce a new dark matter candidate, the lightest $U$-parity particle (LUP).
We suggest it as a good dark matter candidate especially in the $R$-parity violating supersymmetric model.
\end{abstract}

\maketitle


\section{Introduction}
This talk\footnote{The talk was given at SUSY 2008 conference in Seoul, Korea.} is based on several recent papers \cite{Lee:2007fw,Hur:2007ur,Lee:2007qx,Lee:2008pc}.

Supersymmetry (SUSY) needs help from companion symmetries such as $R$-parity and TeV scale $U(1)'$ gauge symmetry to be phenomenologically viable.
A general superpotential of the Minimal Supersymmetric Standard Model (MSSM) is given by
\bea
W &=& \mu H_u H_d \nn \\
&+& y_E H_d L E^c + y_D H_d Q D^c + y_U H_u Q U^c \nn \\
&+& \left( \lambda LLE^c + \lambda' LQD^c + \mu' LH_u + \lambda'' U^cD^cD^c \right) \nn \\
&+& \frac{\eta_1}{M} QQQL + \frac{\eta_2}{M} U^cU^cD^cE^c + \cdots
\eea
where the terms in parentheses are the renormalizable lepton number ($\lnum$) violating terms and baryon number ($\bnum$) violating terms, which are forbidden by the $R$-parity.
As the superpotential shows, the $\lnum$ violation and/or $\bnum$ violation are one of the most general predictions of SUSY.

This superpotential has some problems.
The $R$-parity was introduced to prevent fast proton decay.
It forbids all renormalizable level $\lnum$ violating terms and $\bnum$ violating terms, which is not necessary since forbidding only either $\lnum$ violating terms or $\bnum$ violating terms would be sufficient in preventing proton decay through the renormalizable terms.
But, the $R$-parity makes all of these vanish and precludes some potentially interesting phenomenology associated with these terms.
Furthermore, the $R$-parity allows dimension five $\lnum$ and $\bnum$ violating terms (such as $QQQL$ and $U^cU^cD^cE^c$), which can still mediate too fast proton decay \cite{Weinberg:1981wj}.
Therefore, the $R$-parity by itself is incomplete in addressing the proton stability.
Besides, the superpotential has the so-called $\mu$-problem \cite{Kim:1983dt}, which requires another mechanism or symmetry for the solution.
The issues of the $R$-parity conserving MSSM suggests to look for an additional or alternative explanation.

In this talk, we set our goal to construct a stand-alone $R$-parity violating TeV scale SUSY model without the $\mu$-problem, proton decay problem, and dark matter problem.
We will show that the $R$-parity violating $U(1)'$ model can be an alternative to the usual $R$-parity conserving model.
TeV scale $U(1)'$ gauge symmetry is motivated to solve the $\mu$-problem by replacing the original $\mu$ term ($\mu H_u H_d$) with an effective $\mu$ term ($h S H_u H_d$).
The effective $\mu$ parameter
\be
\mu_{\rm eff} = h \left<S\right> \sim {\cal O}(\mbox{TeV})
\ee
is dynamically generated when the $U(1)'$ is spontaneously broken by the Higgs singlet $S$ with a TeV scale vacuum expectation value (vev). 
See Ref. \cite{Langacker:2008yv} for a recent review of the $U(1)'$ gauge symmetry.
Here we will use the residual discrete symmetry of the $U(1)'$ to address the proton stability and dark matter stability.

\section{Proton stability without $R$-parity}
The most general $Z_N$ discrete symmetry compatible with the MSSM sector is given in Ref. \cite{Ibanez:1991pr}.
\be
Z_N^{vis} :~ B_N^b L_N^\ell
\ee
\begin{figure}[tb]
  \includegraphics[width=.45\textwidth]{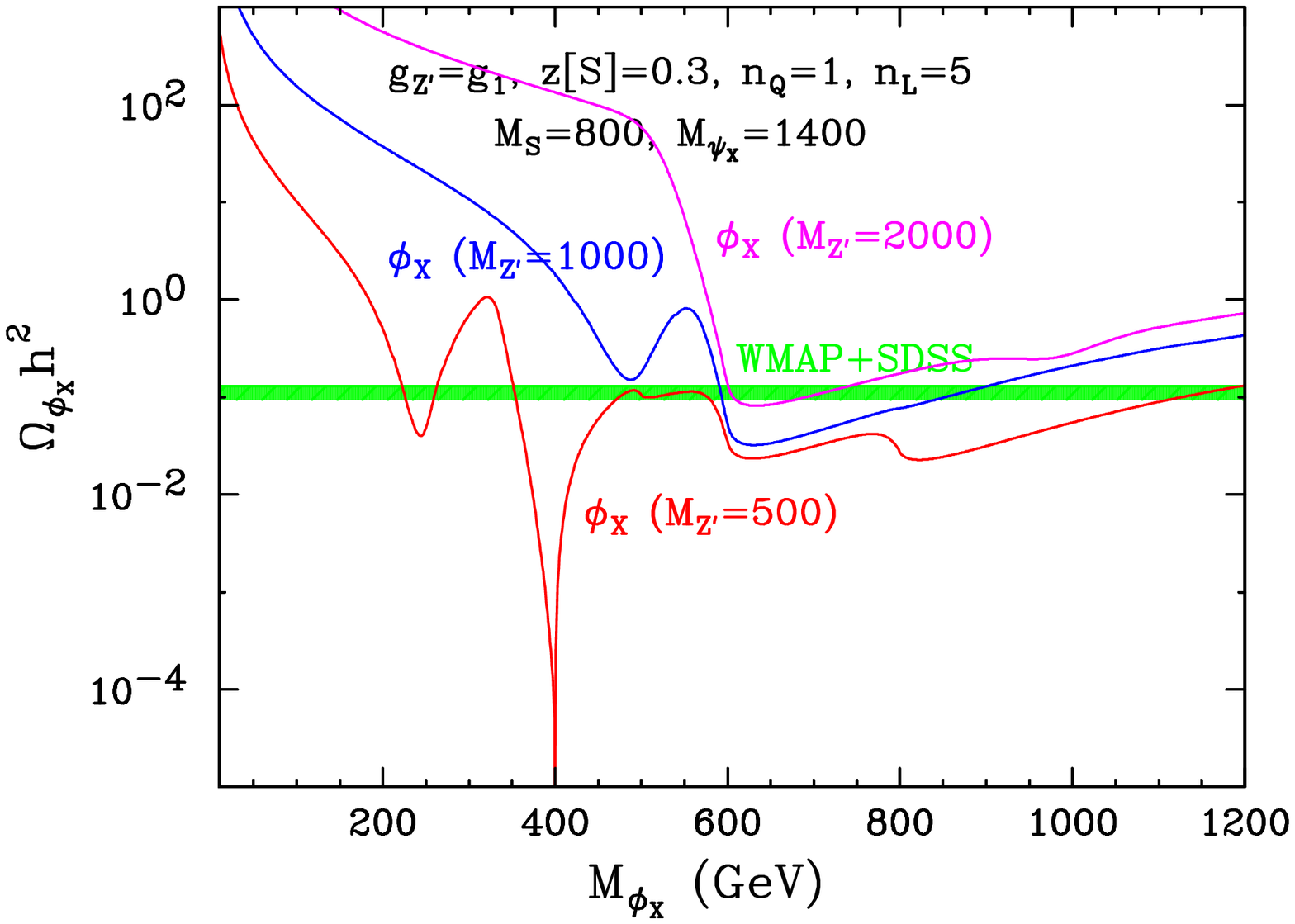}
  ~~~
  \includegraphics[width=.45\textwidth]{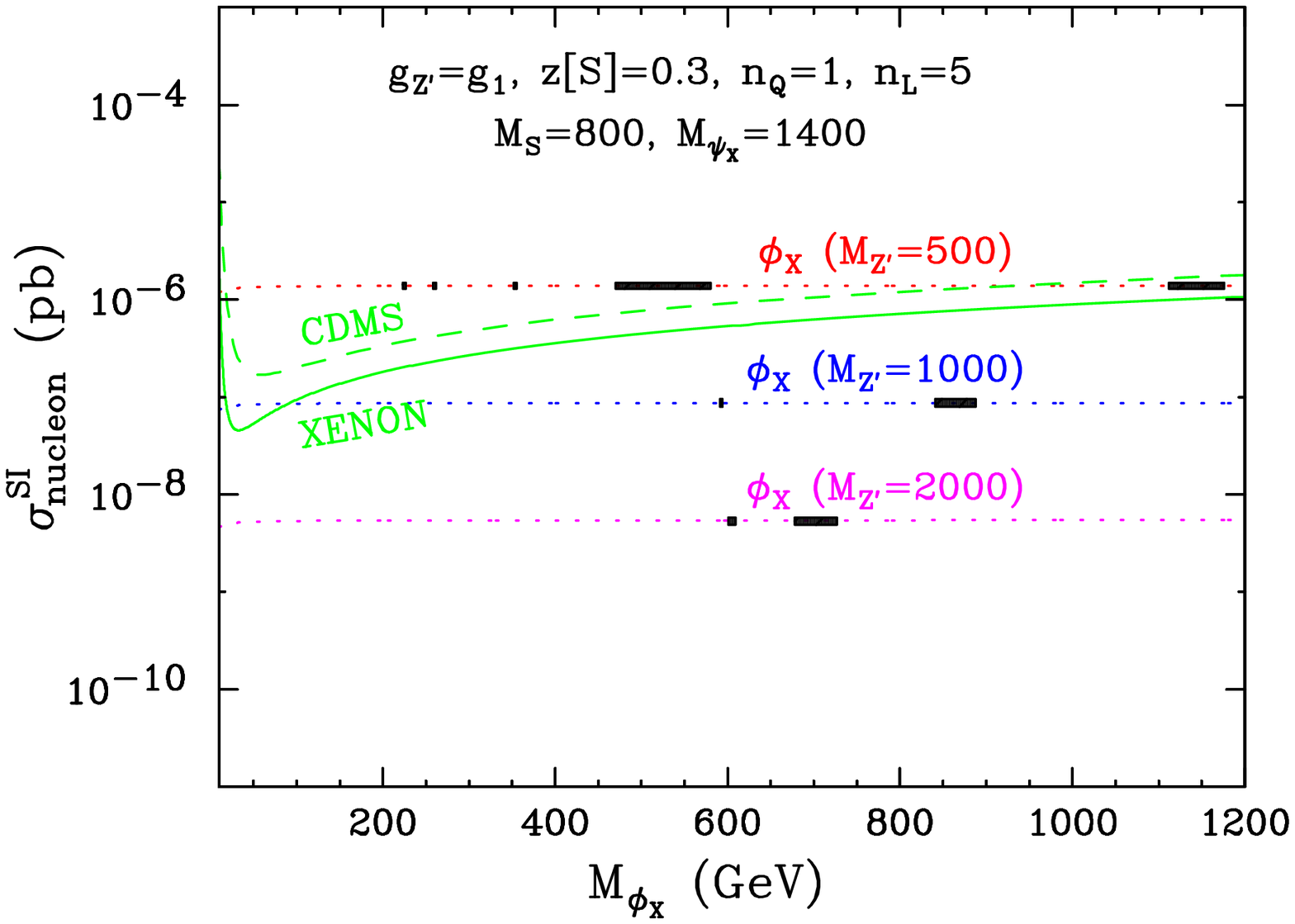}
  \caption{Predictions of relic density and direct detection cross section for a LUP dark matter}
  \label{fig:LUP}
\end{figure}
It was found that the $U(1)'$ symmetry that solves the $\mu$-problem does not allow the simultaneous existence of the $\lnum$ violating terms and the $\bnum$ violating terms \cite{Lee:2007fw}.

We will consider only the $\lnum$ violation case in this talk.
From the superpotential terms and $[SU(2)]_L^2-U(1)'$ anomaly condition, we can have the general $U(1)'$ charges for the MSSM sector for the $\lnum$ violation case.
Then the discrete symmetry can be extracted out of the $U(1)'$ charges, and it turns out it is $B_3$ discrete symmetry, called baryon triality.
In other words, when the $U(1)'$ is assumed to solve the $\mu$-problem, and an effectively renormalizable $\lnum$ violating term (such as $LLE^c$) is assumed, the $B_3$ discrete symmetry is automatically invoked in the MSSM sector as a residual discrete symmetry of the $U(1)'$ \cite{Lee:2007qx}.

Due to the selection rule of the $B_3$
\be
\Delta \bnum = 3 \times \rm{integer} ,
\ee
the proton decay, which is $\Delta \bnum = 1$ process, is completely forbidden under the baryon triality \cite{Castano:1994ec}.

\section{Dark matter candidate without $R$-parity}
Since the $R$-parity is absent, the lightest superparticle (LSP) is not a good dark matter candidate in general.
In principle, it is possible to introduce another symmetry such as global Peccei-Quinn symmetry, which can provide an axion dark matter candidate.
However, we will try to come up with a dark matter candidate in our model without introducing an independent symmetry.
We do that by extending the residual discrete symmetry to the hidden sector, and take the hidden sector field our dark matter candidate.

The SM singlet fields, or hidden sector fields, are often necessary to satisfy the anomaly free conditions with new gauge symmetry, such as $[{\rm gravity}]^2 - U(1)'$ and $[U(1)']^3$.

Their fermionic component can be either Dirac or Majorana, but we will consider only Majorana type for the simplicity, which has the mass term as
\be
W_{\rm hidden} = \frac{\xi}{2} S X X .
\ee
When the $S$ gets a TeV scale vev, the hidden sector field $X$ naturally gets a TeV scale mass.
It is a neutral and massive particle, and it will be good dark matter candidate if it is stable.

Now the question is how to ensure the stability of this particle.
We want to introduce a new parity, which we name $U$-parity \cite{Hur:2007ur}.
Under the $U$-parity, the MSSM fields have even parity, and the hidden sector fields $X$ have odd parity.
\be
U_p[\mbox{MSSM}] = \mbox{even} \qquad U_p[X] = \mbox{odd}
\ee
The lightest $U$-parity particle (LUP) will be the lightest hidden sector field $X$, and it will be stable by the $U$-parity.
Depending on physical masses, the LUP can be either fermionic or scalar component of the $X$.
We do not want to introduce the $U$-parity by hand, and rather we want it a residual discrete symmetry of the $U(1)'$.
\be
Z_N^{hid} :~ U_2 \quad (\mbox{$U$-parity})
\ee
We can achieve it by the charge assignment where the $U(1)'$ charge ($z[F_i]$) is related with the discrete charge ($q[F_i]$) by  
\be
z[F_i] = q[F_i] + 2 n_i
\ee
with the discrete charges $q[\mbox{MSSM}] = 0$, $q[X] = -1$.
Other possible exotics are assumed to be heavier than the lightest $X$ so that they are not protected by the $U$-parity.

\section{Unified picture of stabilities}
Now, how do we have both the $B_3$ and $U$-parity?
We want to consider a $U(1)'$ gauge symmetry with a residual discrete symmetry $Z_6$, which is equivalent to the direct product of the $B_3$ and $U$-parity.
\be
U(1)' ~~\to~~ Z_6^{tot} = B_3 \times U_2
\ee
We can take it as a simple example of a more general case where total discrete symmetry is equivalent to the direct product of the MSSM sector (or observable sector) discrete symmetry and the hidden sector discrete symmetry \cite{Lee:2008pc}.
\be
U(1)' ~~\to~~ Z_N^{tot} = Z_{N_1}^{obs} \times Z_{N_2}^{hid}
\ee
where $N = N_1 N_2$; $N_1$ and $N_2$ are coprime.

\begin{figure}
  \includegraphics[width=.45\textwidth]{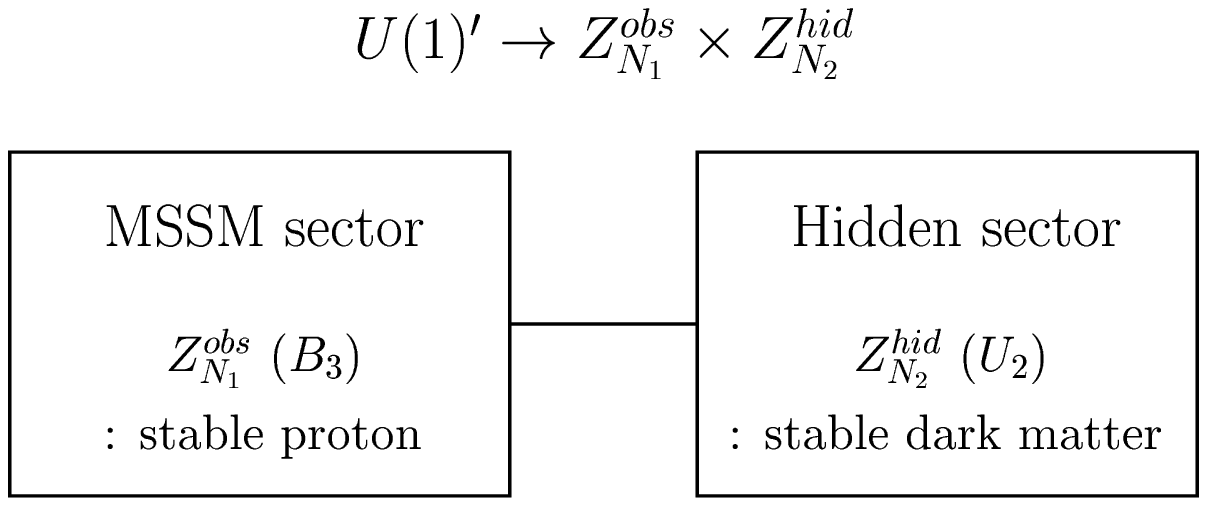}
  \caption{A unified picture of the stabilities in the observable and hidden sectors}
  \label{fig:picture}
\end{figure}
A unified picture about the stabilities in the MSSM sector and the hidden sector arises (Figure \ref{fig:picture}).
A single $U(1)'$ which interacts with both the MSSM sector and the hidden sector provides discrete symmetries for both sectors.
In the usual $R$-parity conserving MSSM, there is a single discrete symmetry which addresses the proton stability and the LSP dark matter stability in the MSSM sector.
Here, there is a single gauge symmetry, which we already have to solve the $\mu$-problem, and it addresses the proton stability in the MSSM sector and the LUP dark matter stability in the hidden sector.

To be a viable dark matter candidate, however, the LUP should satisfy the relic density and direct detection constraints too.
As Figure \ref{fig:LUP} shows, the LUP dark matter can satisfy both constraints simultaneously (see Ref. \cite{Hur:2007ur} for parameter choices).

\section{Summary}
Table \ref{tab:comparison} compares the usual $R$-parity conserving MSSM and the $R$-party violating $U(1)'$ model we considered here.
As it shows, the TeV scale $U(1)'$ gauge symmetry is an attractive alternative to the usual $R$-parity.
\begin{table}[h]
\begin{tabular}{lll}
\hline
               & $R_p$                & $U(1)' \to B_3 \times U_p$~~ \\
\hline
RPV signals    & impossible                & possible \\
$\mu$-problem  & not addressed             & solvable ($U(1)'$) \\
proton         & unstable w/ dim 5 op.   & stable ($B_3$) \\
dark matter    & stable LSP                & stable LUP ($U_p$) \\
\hline
\end{tabular}
\caption{$R$-parity conserving model vs. $R$-parity violating $U(1)'$ model}
\label{tab:comparison}
\end{table}


\begin{theacknowledgments}
This work is supported by the Department of Energy under grant DE-FG02-97ER41029.
\end{theacknowledgments}

\end{document}